# Universal material basis for biocompatible printed electrolytes in Organic Electrochemical Transistors

Moritz Flemming[1*], Paul Zechel[1], Rakesh R. Nair[1], Emil Mahnke[1], Markus Löffler[2], Alyna Ong[2], Bernd Rellinghaus[2], Lukas M. Eng[1,3], Karl Leo[1*], Hans Kleemann[1*]

[1]Institut für Angewandte Physik, TU Dresden, Nöthnitzer Straße 61, Dresden, 01187, Germany.

[2]Dresden Center for Nanoanalysis (DCN), Center for Advancing Electronics Dresden (cfaed), TU Dresden, Georg-Schumann-Straße 11, Dresden, 01069, Germany.

[3]Würzburg-Dresden Cluster of Excellence (EXC2147) ctd.qmat – Complexity, Topology and Dynamics in Quantum Matter, Dresden, Germany

*Corresponding author(s). E-mail(s): moritz.flemming@tu-dresden.de; hans.kleemann1@tu-dresden.de

**Abstract**

Organic Electrochemical Transistors (OECTs) stand out for their interplay between ionic and electronic conduction, making them ideal analogues to biological synapses for neuromorphic computing and biosensing applications. Furthermore, they can be printed into integrated circuits on flexible substrates, enabling low-cost and high-throughput fabrication of complete electronic systems. However, most OECT electrolytes for integrated circuits still lack biocompatibility and suffer from rheology-related printing challenges. This paper presents a novel material basis that can be combined with an ionic liquid to fabricate an electrolyte for OECTs that only contains biocompatible materials. It allows rheological adjustments to enable the use of electrolyte in both inkjet and screen printing. Furthermore, the electrolyte is UV-curable, enabling it to transition into solid-state structures after printing. Extended ink and device lifetimes for screen-printed structures enable the fabrication of advanced OECTs that can operate in ambient air for over 30 days after fabrication. Ultimately, a fully screen-printed transistor using only biocompatible materials on a leaf substrate is shown.

# Introduction

Organic semiconductors that support both ionic and electronic charge transport paved the way for a new generation of organic electronic devices. Among these, a prime example is the Organic Electrochemical Transistor (OECT). The unique working principle of volumetric channel doping allows OECTs to bridge biological interfaces and electronics. Applying a gate voltage forces ions from an electrolyte to enter the device's channel. The channel material belongs to the organic mixed ionic electronic conductors (OMIECs)[1] and allows interaction between the two conduction types. This mechanism allows significant drain current modulations while only low gate voltages are required. This makes them well-suited for applications like ion detection, metabolite sensing, or cell monitoring [2–5]. For sensing in biological environments, a high transconductance and good on/off ratio are desired, since they are directly connected to signal amplification and detection sensitivity. Easy deposition methods and the biocompatibility of many organic materials make OECTs attractive for bioelectronic interfaces and printed biosensors[6].

A central component for OECT operation is the electrolyte, which affects efficiency, speed, and stability of the transduction mechanism. As the (de-)doping of the channel material is driven by ionic interactions, ionic liquids (ILs) are especially interesting. They can be self-dissolving at room temperature and may have negligible vapor pressures, increasing stability. When incorporated into a polymer matrix, ILs form an ionogel that can exhibit higher ionic conductivities than iongels that are purely based on polymers [7,8]. However, many reported gel electrolytes suffer from poor long-term stability: They exhibit moisture loss or evaporation that decreases device reliability. Further, most organic electrolytes lack biocompatibility and therefore have limited application in biological environments.

OECT fabrication simultaneously requires scalable and cost-effective manufacturing methods. Due to its low cost, high throughput, and compatibility with flexible substrates, screen printing is a very attractive deposition technique. It has already been demonstrated to screen print OECTs with a very high yield of over 99%[9]. Nevertheless, developing a screen printable electrolyte that simultaneously provides high shelf life and an extended device stability has remained a challenge. The transition to inkjet printing, which operates in a significantly different viscosity regime, usually requires a separate ink reformulation. Effectively, inks for screen printing and inkjet printing deploy entirely different electrolytes. The possibility to tune a single electrolyte system for both deposition methods without limiting the electrochemical performance or biocompatibility would be a meaningful advancement.

We here present a polymer matrix consisting of biocompatible materials that acts as a universal basis for printable OECT electrolytes by incorporating ionic liquids. By varying the IL type, the rheological and electrochemical properties of the ink can be tuned to allow both screen printing and inkjet printing with the same material basis. The screen-printable formulation has an extended shelf life and shows superior device performance and stability. Combined with a high transconductance, this material basis is well-suited as a foundation for stable, printed OECT biosensors.

# Methods

## Screen Printable Electrolyte

The screen printable electrolyte was prepared by dissolving 200 mg polyvinyl alcohol (PVA, Merck KGaA) in 1 ml dimethyl sulfolxide (DMSO, Thermo Fisher Scientific) and stirring it on a hotplate at a temperature of 100 °C. In a separate vial, 134 µl polyethylene glycol diacrylate (PEGDA, Merck KGaA) was added to 0.6 ml DMSO. Additionally, 50 µl of a 2.5 wt% lithium phenyl-2,4,6-trimethylbenzoylphosphinate (LAP, Merck KGaA) in DMSO solution was injected. Afterward, this mixture was added to the dissolved PVA while maintaining the temperature at 100 °C.

Finally, 1 ml tris(2-hydroxyethyl)methylammonium methylsulfate ([MTEOA][MeOSO$_3$], Merck KGaA) and 14 µl polysorbate 80 (Merck KGaA) were incorporated.

## Device Fabrication

Gold electrodes to serve as the transistor terminals were prepared via photolithography on 1 inch x 1 inch borosilicate glass substrates. The substrates are covered with 3 nm of chromium (Cr) and 50 nm of gold (Au). As a first step, the photoresist AZ 1518 (MicroChemicals GmbH) was spin coated at 3000 rpm for 60 s onto the substrate. Following, annealing for 60 s at 113 °C was carried out. UV exposure of the samples was done with a MJB4 mask aligner at 365 nm for 14.5 s (SÜSS MicroTec SE). The photoresist was then developed in a bath of AZ 726 MIF (MicroChemicals GmbH) for 60 s. Etching of Au was carried out with Standard Gold etchant (Merck KGaA) for 60 s, followed by Cr etching with Standard Chromium Etchant (Merck KGaA) for 15 s. Finally, the samples received O$_2$ plasma cleaning for 10 min.

## Screen-Printed Devices

Channel and gate of the OECT were screen printed with Poly(3,4-ethylene-dioxythiophene) polystyrene sulfonate screen printable ink (PEDOT:PSS, Merck KGaA).

Screen printing was realized with a 120-34 screen at a squeegee movement speed of 50 mm/s, followed by baking at 120 °C for 5 min.

The electrolyte was applied by screen printing at 30 mm/s squeegee speed with a 36-90 screen. Lastly, the structures were hardened by exposing the samples to UV light (365 nm) for 2 min.

## Inkjet-Printed Devices

PEDOT:PSS Clevios PH1000 (Heraeus Deutschland GmbH \& Co. KG) was spin coated at 3000 rpm for 60 s and subsequently baked at 120 °C for 20 min. Next, the photoresist OSCoR 5001 (Orthogonal Inc.) was similarly spin coated at 3000 rpm for 60 s, and annealing was then carried out at 100 °C for 60 s, followed by UV exposure in the mask aligner for 12 s, and a second baking step at 100 °C for 60 s. Next, the resist was twice developed by spin coating the Developer 103a (Orthogonal Inc.) on the substrate. $O_2$ plasma cleaning for 5 min removed the excess amounts of PEDOT:PSS. Resist stripping was achieved by leaving the sample in Stripper 900 (Orthogonal Inc.) overnight. Prior to applying the electrolyte, the surface adhesion of the substrate was increased. Therefore, 9 ml of ethanol, 90 µl of silane A174, and 180 µl of concentrated acetic acid were mixed together. The substrate was immersed in this solution for 15 minutes at 50 °C, then cleaned with a stream of nitrogen, and subsequently baked again for 10 minutes at 100 °C.

The precursor solution for the inkjet-printable electrolyte was prepared by dissolving 15 mg PVA in 1 ml DMSO. The crosslinker solution was prepared in a separate vial similar to the screen-printed mixture. After adding these two solutions together, 0.35 ml of [MTEOA][MeOSO$_3$] was added. Next, 8 drops of polysorbate 80 were injected. The electrolyte was printed onto the substrate using a Fujifilm Dimatix DMP-2800 inkjet printer (Inc. Dimatix Materials). A 12-nozzle "Samba" piezoelectric drop-on-demand (DoD) cartridge was used after the cartridge had been cleaned using a syringe pump. For the printing process, a drop spacing of 15 µm and a nozzle temperature of 42 °C were set. A total of eight layers of the electrolyte were printed onto the substrate. The precursor solution was then exposed to UV light for 120 s to cross-link the electrolyte.

## Current-Voltage Characterizations

Unless stated otherwise, electrical characterizations were conducted in air under standard cleanroom conditions. Measurements were performed using a 4-channel Keithley 4200-SCS Semiconductor Characterization System (Keithley Instruments).

## Impedance Measurements

Impedance measurements were performed on a Metrohm Autolab PGSTAT302N (Metrohm AG).

## Viscosity Measurements

Viscosity measurements were realized with a HAAKE MARS Rheometer (Thermo Fisher Scientific).

## FTIR Measurements

Fourier transform infrared spectroscopy measurements were done with a VERTEX 80v FT-IR spectrometer (Bruker Corporation). For that purpose the precursor mixture was applied to a gold on silicon substrate (Platypus Technologies), and measurements were taken in reflection mode.

# Results and Discussion

## Screen-Printable Electrolyte

The screen printable electrolyte should satisfy three major requirements: A high ionic conductivity, operational and shelf-life stability, and a structure patterning possibility. Due to these demands and its necessary rheological properties, we refer to it as "ionic gel ink" (IGI).

Polyvinyl alcohol (PVA, Fig. 1a) is used as the binding material due to its biocompatibility and capability to serve as an adhesive [10] [11]. Dimethyl sulfoxide (DMSO) serves as the preferred solvent over water because of its low vapor pressure of 5.6 kPa at 20 °C[12]. Mechanical stability and structurability is ensured by the biocompatible crosslinking component polyethylene glycol diacrylate (PEGDA)[13]. PEGDA can polymerize into a hydrogel when appropriate free radicals are provided. This is addressed by the photoinitiator lithium phenyl-2,4,6-trimethylbenzoylphosphinate (LAP) upon exposure to UV light. To improve surface wetting and stability of the dispersion, the surfactant polysorbate 80 is added.

These components form a material basis that allows the incorporation of ionic liquids (ILs) to serve as an electrolyte. It is optimized for ILs that are self-dissolving at room temperature and have a low vapor pressure. In this work, the most commonly deployed ionic liquid is [MTEOA][MeOSO$_3$] due to its biocompatibility and well-studied compatibility with PEDOT:PSS [2]. Unless stated otherwise, the results presented for the IGI are based on the incorporation of this ionic liquid.

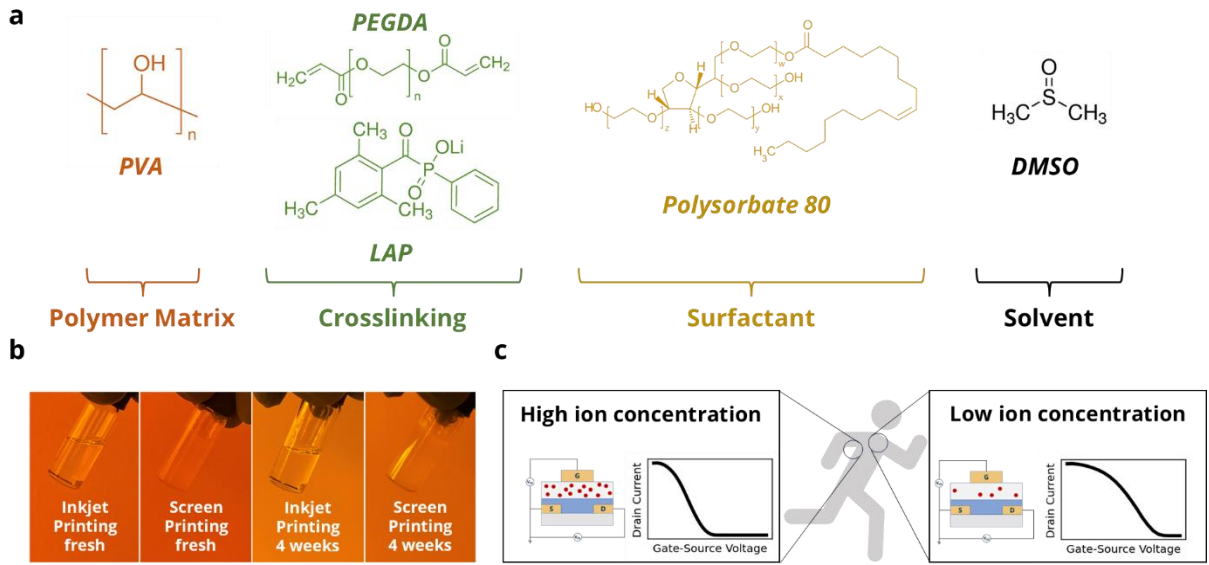

*Figure 1: a) Components used in the material basis. Adding an ionic liquid forms the ionic gel ink (IGI). (b) IGI solutions based on [MTEOA][MeOSO$_3$] for inkjet / screen printing freshly prepared and after 4 weeks without differences visible. (c) Application of an OECT as a sensor to analyze the sweat production for athletes.*

The beforementioned components allow to prepare electrolyte inks for both screen printing and inkjet printing. The inkjet printing process allows the deposition of significantly smaller structures, enabling more compact device layouts. Furthermore, the layer thickness can be precisely controlled, as individual layers can be printed sequentially in a targeted manner. In addition, the inkjet printing process offers greater material efficiency, because it only uses the amount of ink that is specifically printed at the intended locations.

In contrast to screen printing, the ink requirements for inkjet printing are more strictly defined. For screen printing, inks can have viscosities in the range of (1000 – 10000) mPas[14]. According to Shin et al[15], the viscosity of an inkjet-printable ink should, however, be in the range of (1 – 20) mPas. To achieve values in this range, the content of the main viscosity-determining materials is lowered. This especially governs PVA, but may also include the IL, depending on its rheological properties. The composition chosen here represented the best compromise between ink stability and component performance. An increase in IL concentration leads to increased viscosity of the electrolyte, which can be compensated for by further reducing the PVA content to maintain printability in the inkjet printing process. This adjustment results in reduced system impedance. However, reducing the PVA content compromises the mechanical stability of the printed structures as well as their long-term stability during electrical measurements. Ink optimization led to a mixture of 15 mg PVA and 0.35 ml IL that offers the best balance between ink stability and component performance. The overall lowered concentration of ions in the electrolyte is compensated by the higher ionic mobility in a low-viscosity solution. A further difference to screen printing is the drop formation, which requires an appropriate surface tension. This can be achieved by accordingly adjusting the concentration of the surfactant.

The composition and viscosity of the different inks also affects their morphology. Hence, the tilting angle visibly impacts the ink's distribution inside the vial. As shown in Fig. 1b, the lower amount of PVA and IL also results in a higher clarity for the inkjet-printable solution. The IGI solutions are constantly stored under ambient conditions and yellow-light illumination. We cannot observe any major changes in the ink's morphology within 4 weeks after fabrication, which is an indication of a long shelf life.

All the components presented here have been chosen because of their biocompatibility. This is a promising approach to introduce the sensing properties of OECTs into biological environments, i.e., the human body. Depicted in Fig. 1c, a possible application is the sweat monitoring for athletes: The ion concentration in the human sweat will determine the switching behavior of the OECT. Hence, obtaining the electrical characteristics of the OECT allows us to distinguish different concentrations.

## Ink Properties

We first investigate the ink's rheological properties. As the viscosity significantly changes over the stages of the screen printing process[16], a shear rate dependent measurement is performed (Fig. 2a). We found the viscosity of the screen-printed IGI to be at 3500 mPas at a shear rate of 230 $s^{-1}$. The inkjet-printed IGI shows a considerably weaker shear rate sensitivity, its viscosity remains between (10-100) mPas over the whole assessment range. The measurement was again performed 4 weeks later for the same inks and shows there is almost no change in viscosity. For the inkjet-printed solution, the viscosity even slightly decreases, improving its printing behavior. Therefore, our ink has a shelf life that is considerably higher than for other reported screen printable electrolytes, which are typically only demonstrated to be stable for up to 10 days or do not report shelf life at all [9,17–20]. It is worth emphasizing that this shelf life only requires the absence of UV-light and does not need cooling or a nitrogen atmosphere. With this ink formulation, we are able to achieve screen-printed structure widths of down to 200 µm (Fig. 2b). While the usage of higher resolution screens is possible, the printed film thickness is simultaneously reduced, decreasing the switching efficiency. It is hence possible to further increase resolution, but at the cost of deteriorated OECT operation. A SEM image of the screen-printed can be seen in Fig. S1. For inkjet printing, we can go to even higher resolutions of 60 µm width.

A controlled patterning of the printed electrolyte structures is possible due to the combination of crosslinker PEGDA and photoinitiator LAP. Upon UV-light exposure, the free radicals created by the photoinitiator cause the carbon double-bonds of PEGDA to break. Following, the PEGDA monomers can connect into a three-dimensional network that builds a solid hydrogel. To verify this process, the Fourier transform infrared (FTIR) spectra of IGI thin films are investigated. The carbon double bond has a characteristic vibration peak at around 1636 $cm^{-1}$ [21]. For successful PEGDA polymerization, this peak should significantly shrink. We hence compare two IGI thin films: One was exposed to UV-light, the other one did not receive any further treatment (Fig. 2c). We see an almost complete vanishing of the absorbance peak at 1636 cm$^{-1}$. This indicates a successful formation of polymerized PEGDA, sufficient availability of photoinitiator, and good mixing.

Patterning by crosslinking is crucial to ensure the shape of the printed electrolyte structures is preserved. The polymerized PEGDA builds a stabilizing network that improves the mechanical rigidity and stability of the printed structures. Further, the crosslinking networks are increasing in density towards the surface, acting as an evaporation barrier[22–24]. In the device context, it ensures a reduction of parasitic ionic currents and hence efficiently couples gate and channel.

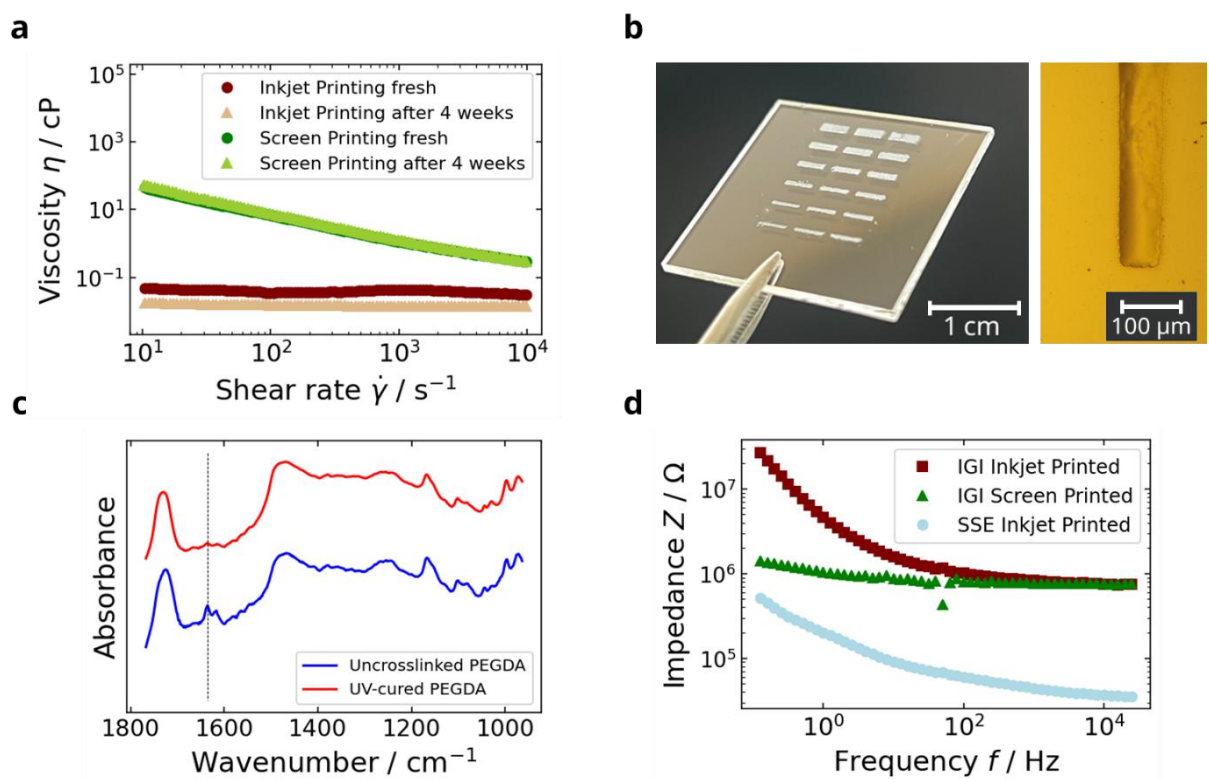

*Figure 2: (a) Shear rate dependent viscosity of IGI formulations. (b) Micrograph of screen printed and inkjet IGI structures. For screen printing, the resolution can be downscaled to 200 µm, while 60 µm are possible for inkjet printing. (c) FTIR spectrum of IGI thin films before and after UV curing. The characteristic C=C double-bond vibration peak at 1636 cm$^{-1}$ almost completely vanishes when PEGDA polymerizes. (d) Impedance of IGI films compared to a solid-state electrolyte (SSE)[25].*

We also look at the electronic behavior of the electrolyte inks. In Fig. 2d, we compare the impedance spectra of both IGI inks to a state-of-the-art solid-state electrolyte (SSE[25]). We observe that both IGI ink formulations have a similar bulk response but differ in their capacitive behavior, which arises due to the difference in ionic mobilities. The ionic conductivity for the screen-printable formulation is determined as 1 mS cm$^{-1}$. The SSE, which deploys [EMIM][EtSO$_4$] as the IL, has an ionic conductivity of 24 mS cm$^{-1}$ [25]. While the values for IGI are around one order of magnitude lower, they lie well within the range of a 0.01 M NaCl aqueous solution[26]. Further differences also govern the IL, as the SSE and IGI are using different ionic liquids which affects conductivity.

## Transistor Characteristics

We used the ionic gel ink to build PEDOT:PSS-based OECTs with gold electrodes. The electrode layouts used for screen-printed devices have a channel length of 240 µm and width of 1200 µm. For inkjet-printed devices, these values are 30 µm and 150 µm, respectively. These transistors are p-type and work in depletion mode: Due to its intrinsic doping, PEDOT:PSS is a hole conductor in its oxidized state. To change this state, a positive voltage between gate and source needs to be applied. Cations from the electrolyte will then migrate into the channel area and compensate for the PSS$^-$ anions. According to the electrochemical reaction

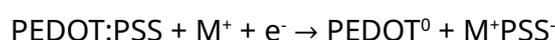

$$\text{PEDOT:PSS} + \text{M}^+ + \text{e}^- \rightarrow \text{PEDOT}^0 + \text{M}^+\text{PSS}^-$$

the PEDOT$^+$ is reduced. Its neutral state, PEDOT$^0$, is not conductive, hence the transistor is switched off. This process is reversible, i.e., upon reduction of the gate voltage, diffusion-driven re-doping of the channel takes place. For the switching-off process, we obtain OECTs with screen-printed IGI that have a threshold voltage $V_{\text{th}} \approx 1$ V (Fig. 3a). State-of-the-art OECTs with inkjet-printed electrolytes can achieve lower threshold voltages[25,27,28]; however, the limiting factor here is the device dimensions constrained by screen printing. The device has a maximum transconductance of $|g_{\text{m}}|$=1.3 mS. To the best of our knowledge, this is substantially higher than previously reported OECTs with screen-printed electrolyte[17,29,30].

In Fig. 3b, the transfer characteristics of this device are compared to OECTs with different IGI formulations. We find an On/Off ratio of 2600 for the screen-printed device, while this value increases by one order of magnitude to 23000 if we use inkjet-printed IGI. We can show the universality of the material basis by incorporating a different ionic liquid: We decided to use [EMIM][EtSO$_4$] as the IL as that is also deployed in the SSE. While lacking biocompatibility, it reportedly has one of the best dedoping effects in PEDOT:PSS-based OECTs [31]. We observe that using this IL in a screen-printed IGI yields comparable transfer characteristics to [MTEOA][MeOSO$_3$] and mainly affects the hysteresis. The integration only requires a proper viscosity tuning and therefore easily allows to switch between ILs.

The output characteristics in Fig. 3c are recorded for a device with inkjet-printed IGI. We can see a clear evolution of linear and saturation regimes with pinch-off point, and the absolute drain current follows the expected trend as the gate-source voltage is increased.

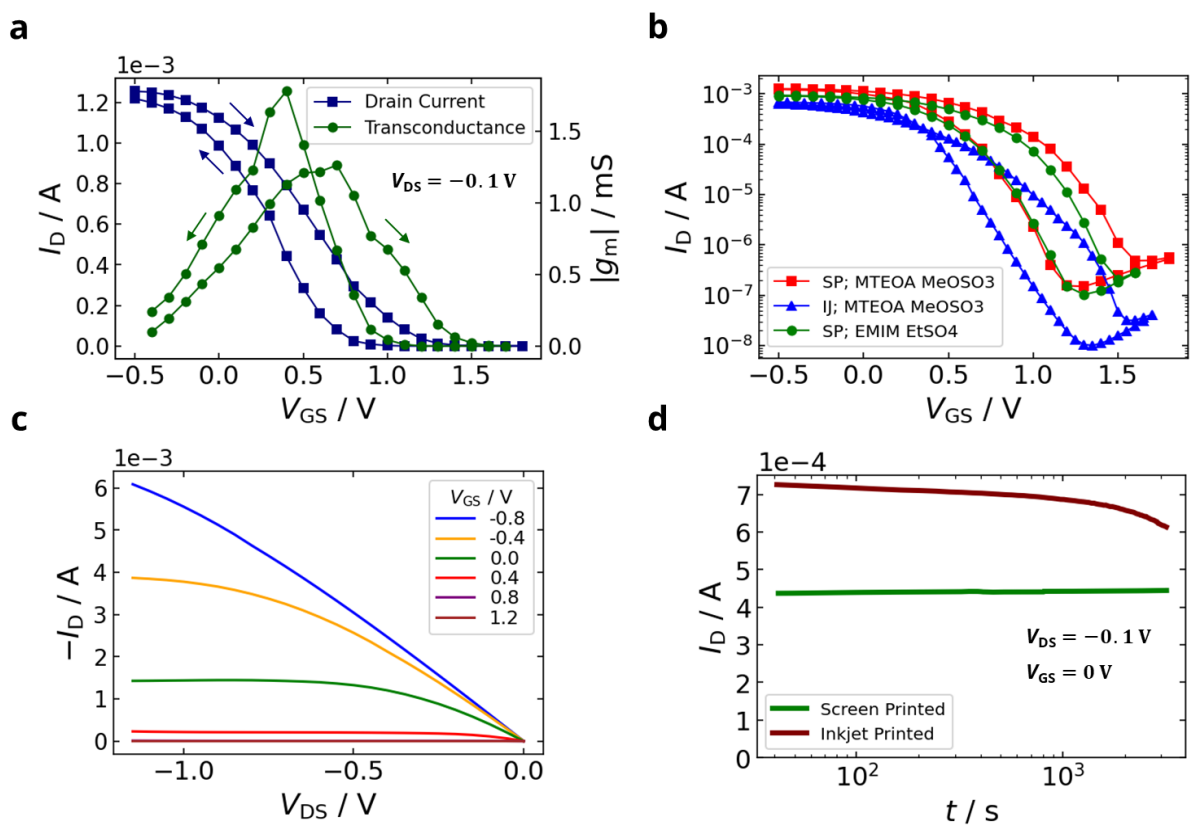

*Figure 3: (a) Transfer curve and transconductance $|g_m|$ of an OECT with screen printed IGI. (b) Transfer curve comparison for three different IGI formulations. (c) Output characteristics of a device with inkjet-printed IGI. (d) Stability analysis for OECTs with IGI under constant gate voltage $V_G = 0$ V.*

A key property that most screen-printed electrolytes lack is stability. In the literature, device lifetimes are rarely reported; for example, Lu et al report values of 10 days[3]. To test stability for our devices, we recorded the drain current for a constant gate-source bias over time. In Fig. 3d, we can see a slight drain current decrease for the inkjet-printed electrolyte, while the screen-printed version remains stable. We attribute this observation to the lower viscosity, i.e., higher ionic mobility of the inkjet-printed electrolyte, making it much more sensitive to external influences. However, the current overall only shows a slight degradation and states good device stability. We further proved cycling stability by consecutively measuring the transfer curves for 80 cycles without any major degradation effects visible (Fig. S2). Additionally, we investigated the long-term stability by measuring an OECT with screen-printed IGI 30 days after fabrication. Without any encapsulation and under ambient storage conditions, the device still shows a similar transfer curve (Fig. S3). Together with the extended shelf life, this material basis allows to fabricate screen printable electrolytes with exceptional stability.

## Biocompatibility

While the components deployed in the IGI have been chosen because of their biocompatibility, applications in biological environments demand every part of the OECT to fulfill this requirement. Given the biocompatibility of PEDOT:PSS[32,33], we are therefore seeking for an alternative to gold electrodes and the glass substrate. The latter is satisfied by lignocellulose-based leaf substrates[34], which are even biodegradable. As these substrates offer flexibility, printing processes are preferred layer deposition methods. We here showcase a fully screen-printed OECT with biocompatible materials: Carbon electrodes, PEDOT:PSS channel, and IGI electrolyte layer are screen-printed onto the leaf substrate.

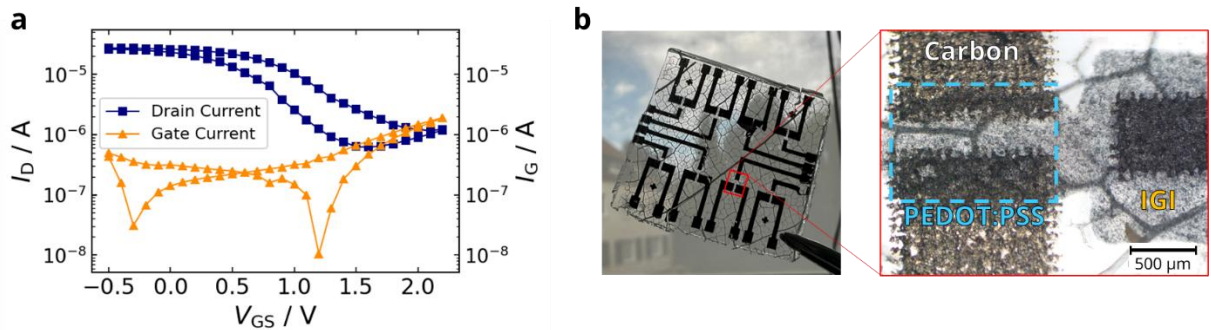

Figure 4: (a) Transfer characteristics of a fully biocompatible OECT on a leaf substrate[34], (b) corresponding substrate / micrograph of the active area.

In Fig. 4b, the substrate with 8 OECTs is shown, including a micrograph of the active area for one of the transistors. The transfer curve in Fig. 4a shows a smaller On/Off ratio and smaller on-current compared to the IGI on regular OECTs (Fig. 3b). We attribute this to the lower conductivity of the carbon electrodes. As visible in the micrograph in Fig. 4b, the screen-printed electrodes also do not form a closed layer. Hence, the voltage drop over the electrodes can be decreased by optimizing the carbon ink and is expected to significantly improve device performance.

# Conclusion

To achieve a stable, screen printable electrolyte for OECTs that only uses biocompatible materials, we developed an ionic gel ink (IGI) that allows universal application for different ionic liquids (ILs) and in different printing techniques. Incorporating the biocompatible IL [MTEOA][MeOSO$_3$] allows us to achieve devices with screen-printed electrolytes that have novel stability. The transistors we can build with the IGI have an exceptional shelf life in ambient air. We showed a high operational current stability and a negligible device degradation over 4 weeks. With transconductance values of 1.3 mS, these transistors pave the way for highly sensitive applications in the human body. We also showed the successful exchange of the IL to [EMIM][EtSO$_4$] with unimpeded device functionality. Adjustments in the ink mixture allow for transition to an inkjet-printable electrolyte. Using the same materials, this allows us to build OECTs with smaller device layouts. Finally, our electrolyte allows us to build fully screen-printed OECTs on biodegradable leaf substrates.

# Supplementary Material

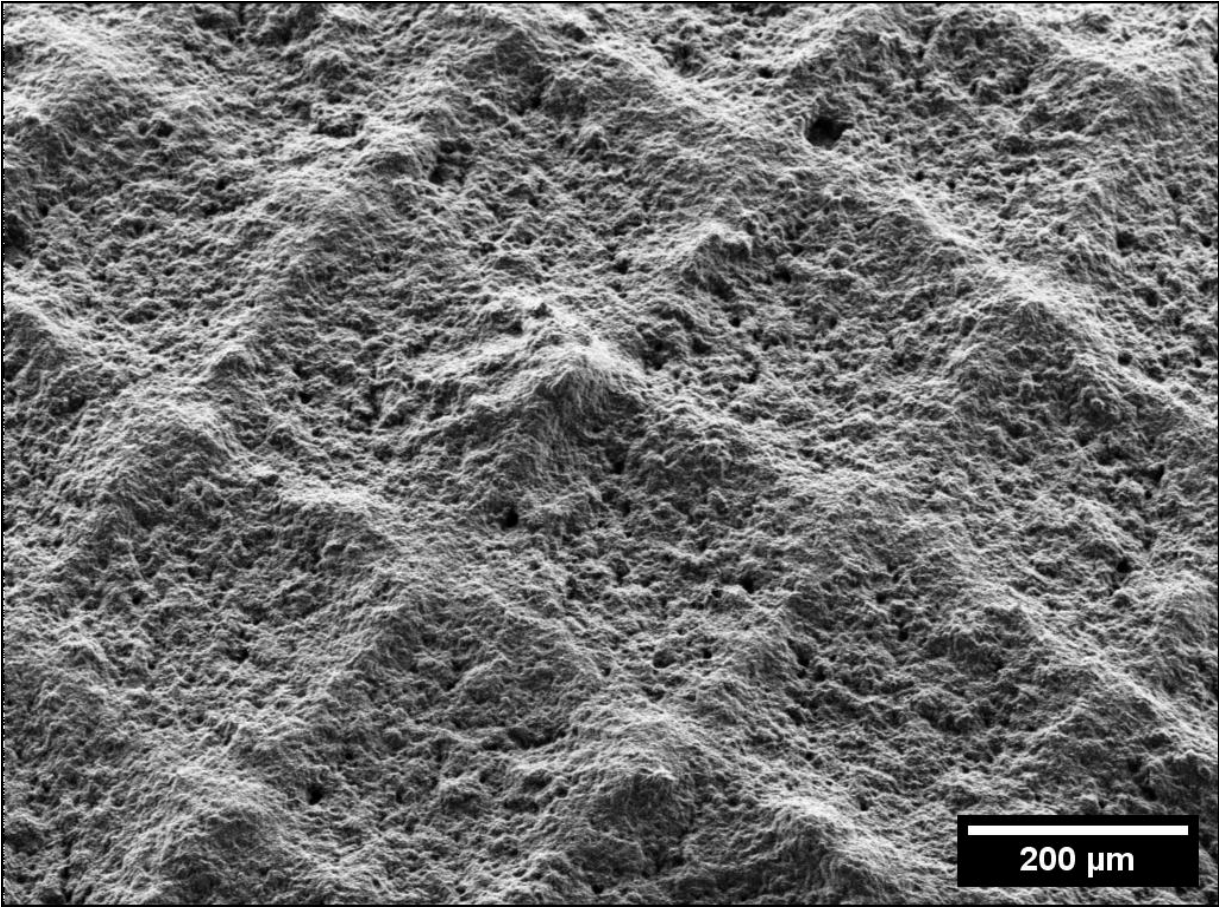

*Figure S1: SEM micrograph of a layer of screen printed IGI with a screen of 36 threads / cm. For better visibility of the topography, the sample was imaged at a tilt angle of 30°.*

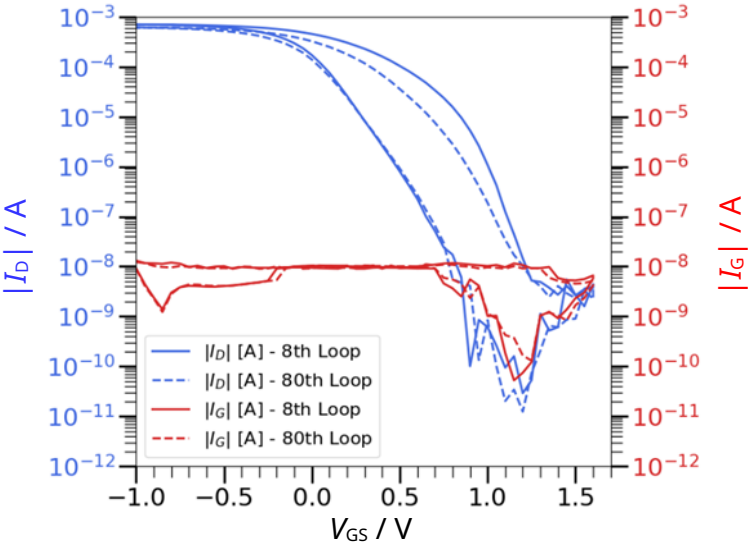

*Figure S2: Device cycling analysis of an OECT with inkjet-printed IGI. Measurement conducted under nitrogen atmosphere.*

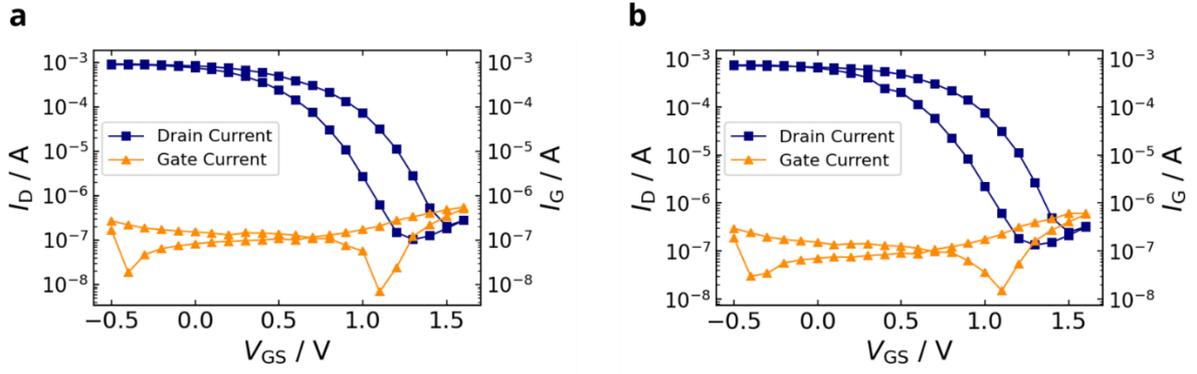

*Figure S3: Transfer curves of OECTs with screen-printed IGI based on [EMIM][EtSO4]. (a) Freshly prepared device, (b) device that has been measured 30 days after preparation.*